\documentclass[prb,twocolumn,showpacs,preprintnumbers,amsmath,amssymb]{revtex4}

\usepackage{graphicx}
\usepackage{dcolumn}
\usepackage{bm}
\usepackage{subfigure}

\usepackage{url}

\begin{document}

\title{Growth and structure of Ag on bilayer Co nanoislands on Cu(111)}

\author{Jakob Bork}
\author{Jens Onsgaard}
\author{Lars Diekh\"{o}ner}%
\email{ld@nano.aau.dk}
\affiliation{%
Institut for Fysik og Nanoteknologi, and Interdisciplinary Nanoscience Center (iNANO), Aalborg Universitet, Skjernvej 4A, DK--9220 Aalborg, Denmark
}%


\begin{abstract}
We have studied the growth of Ag on bilayer high Co nanoislands on
Cu(111) using Scanning Tunneling Microscopy. Noble metal capping of magnetic nanostructures is known to
influence the magnetism and knowledge of the growth is
therefore important. We find that Ag preferentially nucleates on the
Co nanoislands initially leaving the free Cu sites clean.
Furthermore we observe, that those Co islands which are capped with
Ag are almost completely capped thus making a perfect multilayered system
of Ag/Co/Cu(111). We observe a (9$\times$9) reconstruction of the Ag overlayer on
Co/Cu(111).
\end{abstract}


\maketitle

\section{\label{sec:introduction}Introduction}
The properties of metallic thin films and nanostructures on crystalline surfaces are strongly interlinked with the morphology and chemical composition of the structures. Epitaxial growth of in particular homoepitaxial and bimetallic systems have therefore been studied extensively \cite{Venables1984,Brune1998,Ratsch2003}. When growing multilayered heterostructures of several materials it becomes more complicated to control the growth and ensure that the desired structures are fabricated during deposition of the constituents of interest. Besides diffusion on terraces also interlayer mass transport becomes decisive for the growth \cite{Tersoff1994,Bromann1995,Zhang1997,Rottler1999,Krug2000}. Multilayered systems of magnetic and non-magnetic metallic layers, which we have studied, are in particular interesting for their technological applications in the area of data
storage and processing \cite{Prinz1998,Wolf2001,Zutic2004}. It has recently been shown that capping a ferromagnetic thin film with a
noble metal can influence the magnetic
properties \cite{Weber1995, Tsay2001, Kisielewski2002, Chen2002,
Chen2008,Gabaly2008}.

We report here on the capping of ferromagnetic Co nanoislands with Ag thus producing Ag/Co/Cu(111) nanostructures. The growth of Co on Cu(111) is well known and the
structural, electronic and magnetic properties have been studied in
great detail \cite{Figuera1993, Figuera1994, Heinz1994,
Alkemper1994, Figuera1996, Diekhoener2003, Pietzsch2004,
Negulyaev2008}. Important for the work presented here is that for lower coverages at
room temperature Co forms bilayer high islands of triangular
shape and a width of 3--25 nm \cite{Figuera1993, Negulyaev2008}. We have deposited Ag on top and found that there is a
preference for covering the Co islands before the Cu-terraces
and we observe that the Ag capped Co islands, with few exceptions, are always fully
covered with Ag. We conclude that interlayer mass transport plays an important role where Ag atoms, which have landed on the Cu-terrace, must be allowed to ascend the bilayer high step edges of the Co-island.

\section{\label{sec:experiment}Experiment}
Experiments have been performed in an ultra high vacuum system
($5\times10^{-11}$ mbar) using a scanning
tunneling microscope (STM) operated at room temperature and low energy electron diffraction (LEED). We have used WSxM to
analyze and display STM images \cite{horcas:013705}. A Cu(111) crystal was cleaned by sputter/anneal (500$^\circ$C) cycles. Co was
deposited with a flux of 0.14 ML/min at $T$=27$^\circ$C . Next, Ag was deposited
at room temperature (or 70$^\circ$C in one case) with a flux of 0.44 ML/min or
3.6 ML/min. The area Co coverage in all experiments is 0.15--0.36 ML and the Ag overlayer coverage is varied from 0 to $\sim$1.6 ML. Note that for the Co islands the term monolayer (ML) is used for the visible area coverage, meaning that the deposited amount of Co is twice the visible coverage since the Co islands are of bilayer height.

\section{\label{sec:data}Results and Discussion}
\subsection{\label{sec:stm}Growth}
\begin{figure}
\centering
\includegraphics[width=0.4\textwidth]{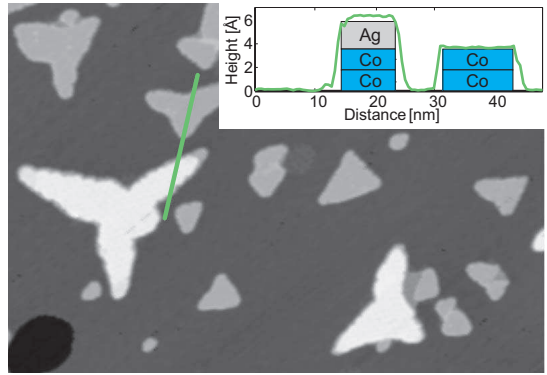}
\caption{\label{fig:20080826m7}
STM image after depositing 0.07 ML Ag on 0.2 ML Co/Cu(111).
The image is taken at a bias voltage of 0.35 V, the size is 157$\times$108 nm and the z-scale spans 0.91 nm.
The inset shows a linescan.}
\end{figure}
\begin{figure}
\centering
\begin{tabular}{cc}
\includegraphics[width=0.4\textwidth]{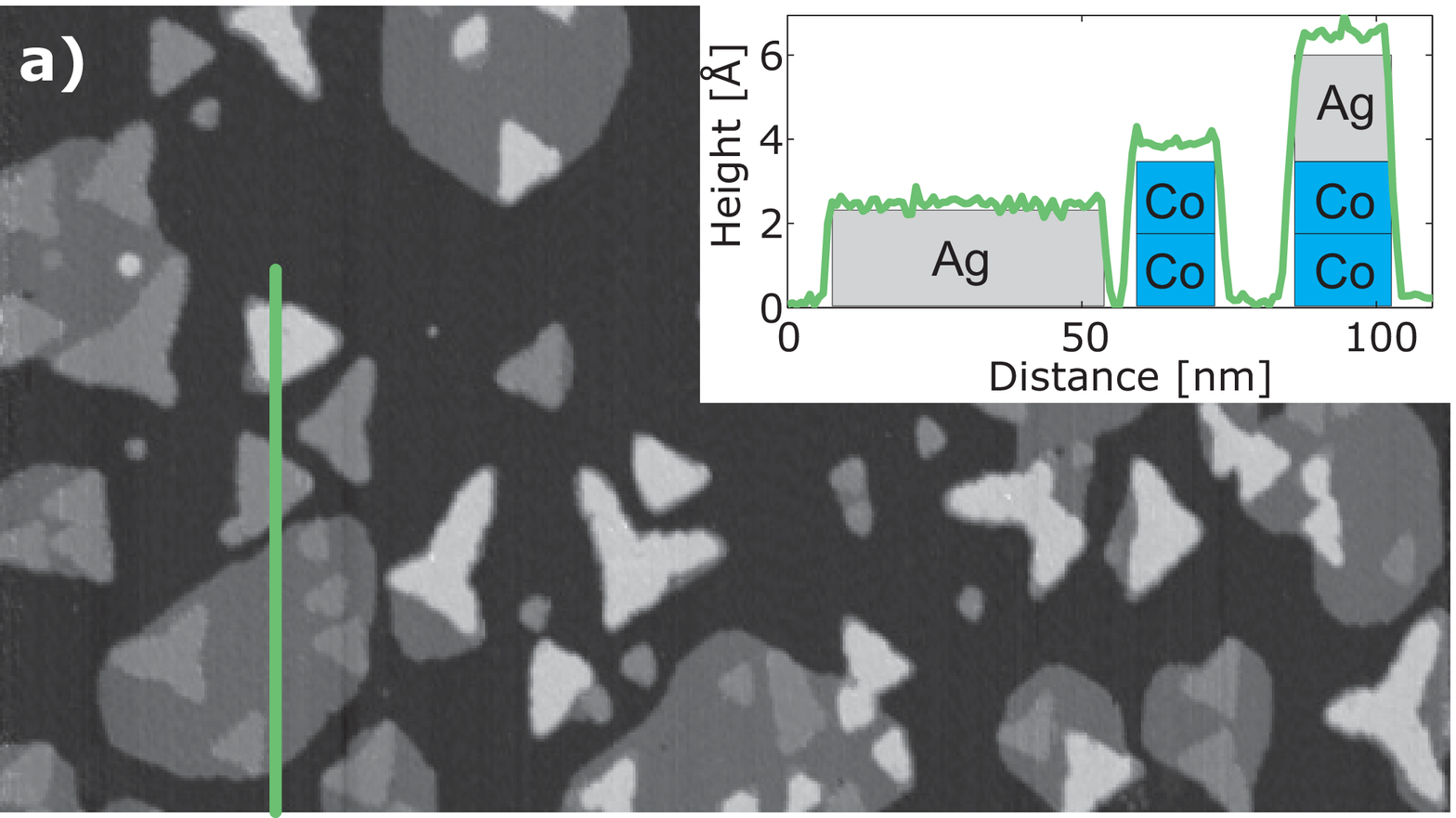}\\
\includegraphics[width=0.4\textwidth]{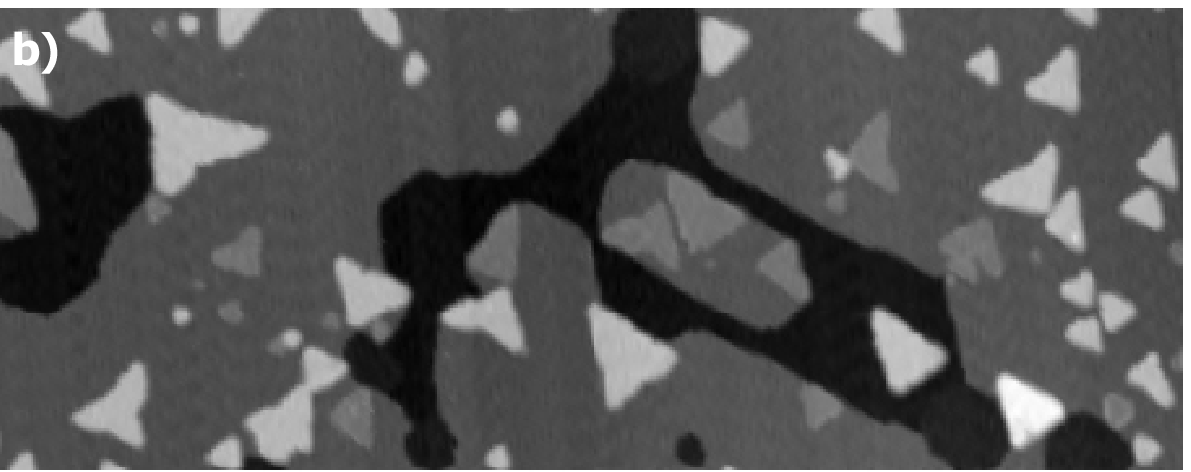}\\
~~\includegraphics[width=0.4\textwidth]{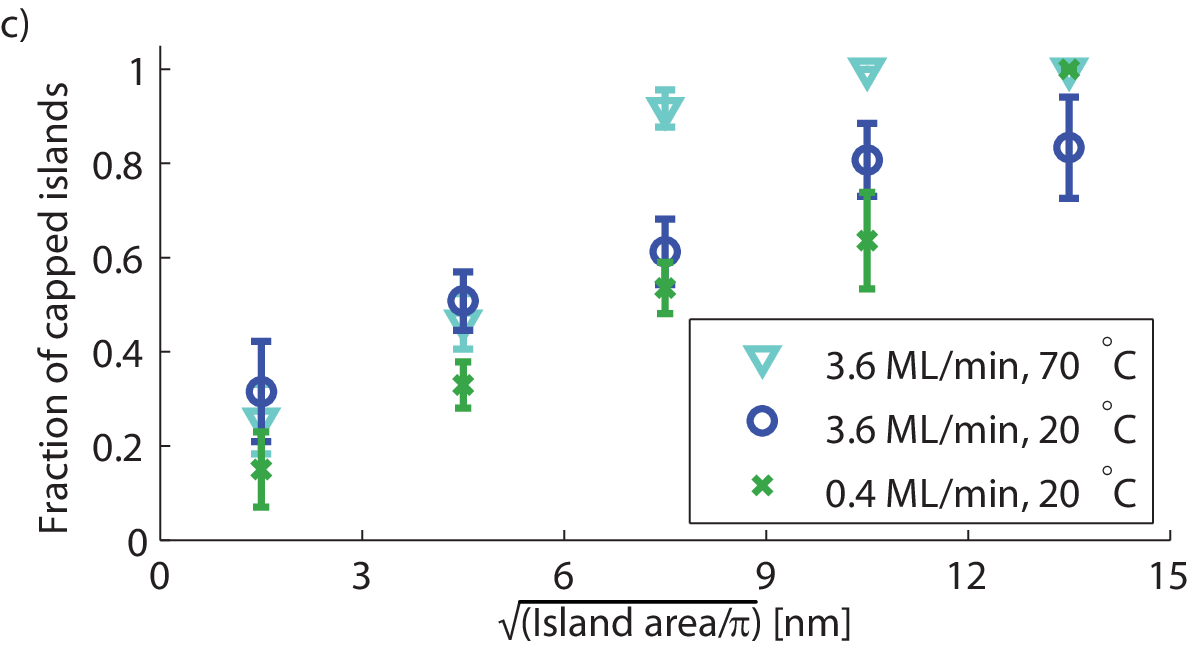}
\end{tabular}
\caption{\label{fig:AgCoCu}(Color online) STM images for
a) 0.35 ML Ag on 0.2 ML Co/Cu(111). The image is
300$\times$167 nm and the z-scale spans 0.94 nm. The linescan runs through a Ag capped Co island, an uncapped Co island and a layer of Ag directly on Cu(111).
b) 0.80 ML Ag on 0.2 ML Co/Cu(111). The image is
300$\times$118 nm and the z-scale spans 0.94 nm. c) The fraction of capped islands as a function of approximate island radius grouped in 3 nm bins for two Ag-deposition rates and two temperatures. $\Theta_{Co}\approx0.17$ and $\Theta_{Ag}\approx0.4$. In total 598 islands were analyzed to produce the 3 data sets.}
\end{figure}

After depositing Co on Cu(111) we observe the well-known bilayer high (3.9
\AA) islands \cite{Figuera1993,Figuera1994,Figuera1996,Diekhoener2003, Pietzsch2004,Negulyaev2008} mostly of triangular shape as
well as star-shaped islands (which is known to be a kinetic effect \cite{Negulyaev2008}). An STM image of 0.07 ML Ag deposited on a Cu surface with a low coverage of Co islands ($\Theta_{Co} = 0.20$, where only bilayer high Co islands are formed) is shown in Figure \ref{fig:20080826m7}. The bright islands are consisting of 2 ML Co with 1 ML Ag on top. Although Co islands of 3 ML height are never observed at these Co coverages we note for comparison that the height of islands consisting of 2 ML Co + 1 ML Ag is 1 {\AA} larger than 3 ML Co. Furthermore the Ag cap layer shows a reconstruction (discussed later). The
less bright islands are 2 ML Co islands that are not capped with Ag.
The heights are illustrated by the inset in Figure
\ref{fig:20080826m7}. It is remarkable that those Co islands which \emph{do contain} Ag on
top are almost \emph{completely covered} with a Ag layer.  On the other hand, the amount of Ag on the Cu surface is minimal although Ag has been deposited homogeneously. Ag thus accumulates
on top of some of the Co islands leaving the other Co islands and
the Cu terrace clean. We note that occasionally it might happen that narrow corners of islands are not filled with Ag (as observable at the upper right corner of the largest star shaped island in Figure \ref{fig:20080826m7}). As the Ag coverage is increased to 0.35 ML we observe, besides fully capped Co
islands, that some of the Co islands are surrounded by Ag instead (Figure~\ref{fig:AgCoCu}a).
At 0.8 ML Ag (Figure \ref{fig:AgCoCu}b) most of the Cu surface and all
Co islands are filled with Ag, except from some of the smaller Co islands. In Figure~\ref{fig:AgCoCu}c we have plotted the fraction of Co islands that are capped with Ag as a function of island size and see that larger islands have a higher probability of being capped.

\begin{figure}
\centering
\includegraphics[width=0.4\textwidth]{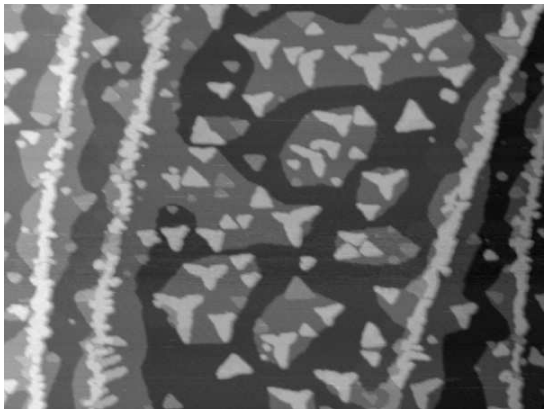}
\caption{\label{fig:overview}Overview STM image of 0.48 ML Ag
deposited on 0.25 ML Co on Cu(111). For the analysis only terrace
sites are used and the areas around Cu-steps are neglected. The
image size is 400$\times$300 nm.}
\end{figure}

Before analyzing the data in more detail we will briefly
discuss the growth in the areas containing Cu step edges. Co
deposition on clean Cu(111) leads to islands on the terraces as
well as a decoration of both sides of the Cu step
edges \cite{Figuera1993}. On the other hand, deposition of Ag on
clean Cu(111) at room temperature never leads to nucleation on the terraces but only to
decoration of the lower side of Cu step edges resulting in "step
flow growth" \cite{McMahon1992}. In agreement with this we find here
that Ag deposited on a partially Co covered Cu(111) surface
nucleates at step edges but also on top of Co islands. This is shown in
an overview STM image in Figure~\ref{fig:overview}. We avoid the
complexity at the Cu step edges and concentrate our further analysis on
Cu terraces.

\begin{figure}
\centering
\includegraphics[width=0.45\textwidth]{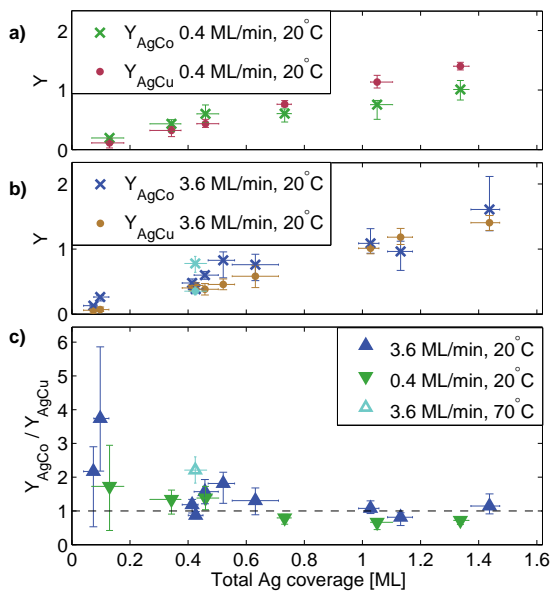}
\caption{\label{fig:relation}(Color online) a) The partial coverage (Y) of Ag on Co and Cu respectively at low Ag-deposition rate (0.44 ML/min).
b) The partial coverage (Y) of Ag on Co and Cu respectively at high Ag-deposition rate (3.6 ML/min).
c) The ratio of the partial coverages of Ag on Co-islands and Ag on Cu(111) as a function
of deposited Ag for the two Ag-rates. At a ratio above 1, indicated by the horizontal dashed line, there is preference for Ag-nucleation on Co. All data are for deposition at $T$=20$^\circ$C except for the light-blue data point ($T$=70$^\circ$C).}
\end{figure}

We will now quantify the preference of Ag nucleating on top of Co
islands vs. nucleating on Cu. For fabrication of
Ag/Co/Cu structures nucleation on top of Co islands would be
preferred. We have determined the partial coverage of Ag on Co islands,
$Y_{AgCo}$, and Ag directly on Cu(111), $Y_{AgCu}$, from STM images
of surfaces with variable Ag coverage and a Co coverage of 0.25$\pm$0.1 ML. The partial coverage of Ag on Co (Cu) is defined as the
relative area of Co (Cu), that is covered with Ag. In this way we
normalize the Ag coverage to the available area of Co (Cu). Whereas $Y_{AgCo}$ is sensitive to $\Theta_{Co}$ it is notable that the free Cu area plays a less important role for how much of the Ag nucleates on Cu sites since nucleation of Ag never occurs on Cu terrace sites
away from Co island steps. We have used the Flooding function of the WSxM program \cite{horcas:013705} to determine coverages.
The images used for the analysis are sized
approximately 200 nm $\times$ 200 nm. For each datapoint in Figure
\ref{fig:relation} we have analyzed 1--4 different areas on the
surface for a given preparation. In Figure \ref{fig:relation}a and \ref{fig:relation}b we plot $Y_{AgCo}$ and $Y_{AgCu}$ as a function of $\Theta_{Ag}$ for two different Ag deposition fluxes of 0.44 ML/min and 3.6 ML/min, respectively. At low Ag doses $Y_{AgCo}$ increases faster than $Y_{AgCu}$. This preference disappears above $\Theta_{Ag}\sim0.7$ML. The effect is more clearly demonstrated by plotting the ratio $\frac{Y_{AgCo}}{Y_{AgCu}}$ (Figure~\ref{fig:relation}c), which thus is a measure of the preference of Ag nucleating on Co vs. on Cu\footnote{Although the absolute errorbars in Figure \ref{fig:relation}a and b are comparable, the relative errors are larger for low Ag coverage. This explains the large errorbars seen at low coverage of the ratio between $Y_{AgCo}$ and $Y_{AgCu}$ in Figure \ref{fig:relation}c}. In agreement with the
qualitative discussion earlier we observe indeed that at low Ag
coverage there is an increased preference of Ag nucleating on Co islands. At
higher Ag coverage this decreases since Co islands which are already capped and those with a Ag rim are no longer available and the ratio approaches unity. Furthermore we note that a higher Ag flux initially leads to a slightly increased preference for nucleation on Co (see Figure
~\ref{fig:relation}c).
We observe no significant dependance on $\Theta_{Co}$ in the range we have used ($\Theta_{Co}$: 0.15--0.36 ML).

\begin{figure}
\centering
\includegraphics[width=0.45\textwidth]{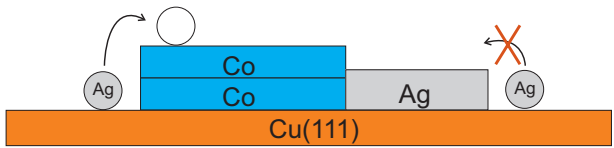}
\caption{\label{fig:climbing}(Color online) We here illustrate the peculiar growth, where Ag atoms ascend bilayer high Co islands but do not ascend monolayer high Ag islands (or Cu step edges).}
\end{figure}

There have been a number of studies investigating nucleation on a surface with pre-grown islands \cite{Tersoff1994,Bromann1995,Zhang1997,Rottler1999,Krug2000}. Atoms landing directly on an island will diffuse around until they either descend and disappear from the island or meet other adatoms finally forming a stable cluster. This can then act as nucleation site for further adatoms. The additional Erlich-Schwoebel barrier at the step edge \cite{Ehrlich1966,Schwoebel1964} increases the time an adatom stays on the island and thereby the probability of meeting another adatom. With this in mind we can rationalize our experimental findings. We observed that larger Co islands have a larger probability of being capped with Ag (see Figure~\ref{fig:AgCoCu}c). This has also been seen in other systems \cite{Bromann1995} and is due to a larger probability of finding two adatoms at the same time on a larger island. There was a more sharp transition between capped an uncapped islands, compared to the results presented here, and it was possible to identify a so-called critical radius ($R_{c}$), where islands having a radius larger than $R_{c}$ would experience 2.~layer nucleation on top \cite{Tersoff1994,Bromann1995}. The varying and in part irregular shape of the Co islands makes it difficult to identify a radius but we used $\sqrt{A/\pi}$, where $A$ is the island area\footnote{The geometry of the step in general influences diffusion barriers \cite{Negulyaev2008} and the island shape may therefore play a role in the interlayer mass transport. It was therefore necessary to use relatively large bin sizes to achieve reasonable statistics and could explain why the capping fraction vs. island size in Figure~\ref{fig:AgCoCu}c is so broad compared to the more step like distribution seen e.g. in Ref \cite{Bromann1995}, where the islands varied in size but where much more uniformly shaped. We note that the size distribution of the Co islands ranges from 3-25 nm in width showing a smooth and broad distribution, peaking at ca 10 nm width. }. We found that a higher Ag flux lead to increased nucleation on top of Co islands. This is seen as a slight shift towards smaller islands in Figure~\ref{fig:AgCoCu}c but more clearly in Figure~\ref{fig:relation}c, where the trend is that the ratio $\frac{Y_{AgCo}}{Y_{AgCu}}$ (and thereby the preference for nucleation on Co) is largest for the high Ag deposition rate, 3.6 ML/min. A higher Ag-flux makes it more likely that another atom necessary for making a stable cluster lands on the island in time. We also studied the effect of increased surface temperature. The simple expectation would be that a higher temperature leads to a higher diffusivity of Ag adatoms on the island and a higher descend rate thus reducing the residence time and the chance of nucleation \cite{Tersoff1994,Bromann1995} (especially visible in the experimental data presented in Ref \cite{Bromann1995}). We find on the contrary that $T$=70$^\circ$C results in an \emph{increased} nucleation rate on the islands compared to $T$=20$^\circ$C. This can be seen in Figure~\ref{fig:AgCoCu}c (as a function of island size) and in the ratio of relative coverages (Figure~\ref{fig:relation}c). The explanation presented in Ref \cite{Tersoff1994,Bromann1995} does not take into account that atoms landing on the bare surface (not on the islands) may \emph{ascend} the islands and thereby lead to an increased nucleation probability on the islands. Apparently this is a prerequisite in this case and is very surprising since the Ag atoms need to ascend steps of 2 ML height. In contrast, Ag atoms never ascend monatomic Cu steps \cite{McMahon1992} or steps at Ag islands. This peculiarity is illustrated in Figure~\ref{fig:climbing}. The reason is likely to be found in Co being chemically more reactive than the more noble Cu \cite{Larsen1998} and the binding of Ag is therefore expected to be stronger to Co than to Cu. This could give rise to asymmetric rates of step edge crossing, i.e.~the diffusion rate for atoms moving up is higher than for moving down, but also to a smaller critical cluster size on Co than on Cu. Both effects lead to a preference for nucleation and an accumulation on Co compared to Cu as observed.

Another reason for the preference of capping Co with Ag can possibly be found in the gain of free energy \cite{Venables1984}. Most simply there are three free energy parameters to take into account, where $\gamma_{S}$ is the surface free energy of the substrate (Cu or Co), $\gamma_{D}$ is the surface free energy of the deposit (Ag in this case) and $\gamma_{i}$ is the interface free energy. The values for the surface energies are $\gamma_{Ag(111)}=1.17~J/m^2$, $\gamma_{Cu(111)}=1.95~J/m^2$ and $\gamma_{Co(0001)}=2.78~J/m^2$ \cite{Vitos1998}. Note that the latter value is for Co(0001) and not for 2 ML Co/Cu(111), where it may be different due to the strained Co adlayer, which has adapted to the Cu lattice. The criterion for layer-growth, $\gamma_{D}+\gamma_{i}<\gamma_{S}$ seems to be fulfilled for both Ag on Cu(111) and Ag on Co/Cu(111) since in both cases a complete adlayer of Ag is formed. The change in free energy, $\Delta\gamma=(\gamma_{D}+\gamma_{i})-\gamma_{S}$ is difficult to evaluate exactly since, to our knowledge, $\gamma_{i,Ag/Co}$ is not available for Ag on Co/Cu(111). The interface energy is strongly interlinked with the lattice mismatch and there is therefore reason to believe that $\gamma_{i,Ag/Co}$ and $\gamma_{i,Ag/Cu}$ are similar in this particular case and the gain in free energy will therefore be larger for capping the Co islands, since $\gamma_{Cu}<\gamma_{Co}$. This argument is in agreement with our finding of increased preference for capping Co.

Although the mobility of single atom diffusion on (111) surfaces is
very high it does not always lead to Ag-covered Co islands. This can be explained by an apparent high critical cluster size for nucleation, i.e.~only if a sufficient number of Ag atoms meet and form a stable
immobile cluster on top of the Co island the Ag adlayer will be
formed. Otherwise, Ag atoms diffuse down on to the Cu surface. If, on
the other hand, a stable cluster has been formed, attachment of
further Ag atoms occurs frequently due to the high mobility of Ag
atoms and a complete adlayer is formed. At very low Ag coverages it might happen that there is not sufficient Ag to form a complete capping layer on an island, but the Ag coverages used here (all above 0.07 ML) lead to either capped or empty Co islands. This is in contrast to findings for Au or Cu growth on Co/Cu(111), where only part of a Co island is capped \cite{Diekhoener2009}. Those Co-islands which are not capped are mostly completely surrounded with Ag (see Figures~\ref{fig:AgCoCu}a and ~\ref{fig:AgCoCu}b). This can be due to a simple blocking effect, where a Ag-rim at the Co-island border prevents further atoms from ascending the island or that the rim reduces the Erlich-Schwoebel barrier and the confinement of Ag atoms landing on top of an island is then less effective.
It is known that Co islands on Cu(111) grow in two general orientations \cite{Figuera1993}. We note that this had no influence on the Ag-capping.

\subsection{\label{sec:leed}Structure}
\begin{figure}
\centering
\begin{tabular}{cc}
\includegraphics[width=0.2\textwidth]{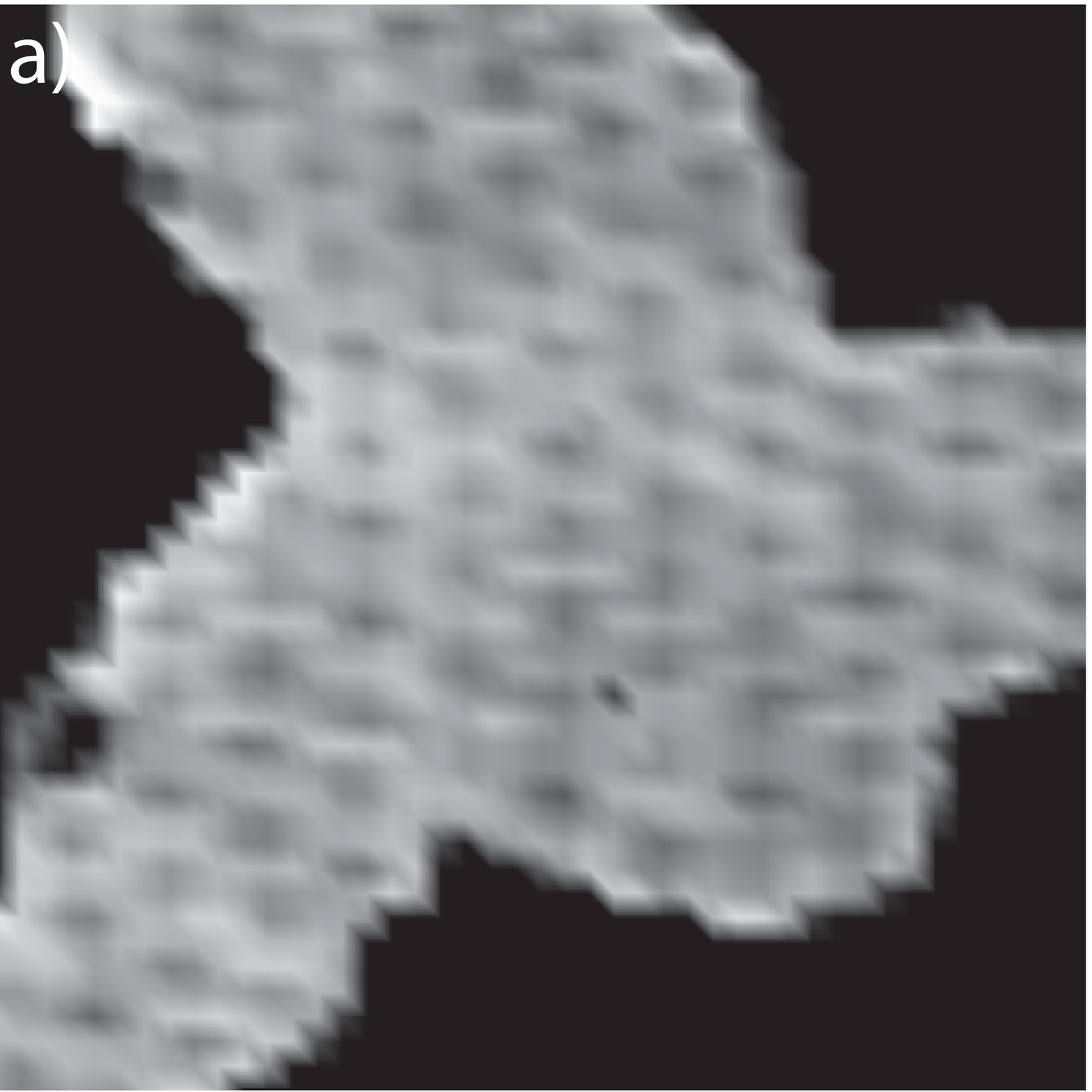}&\includegraphics[width=0.2\textwidth]{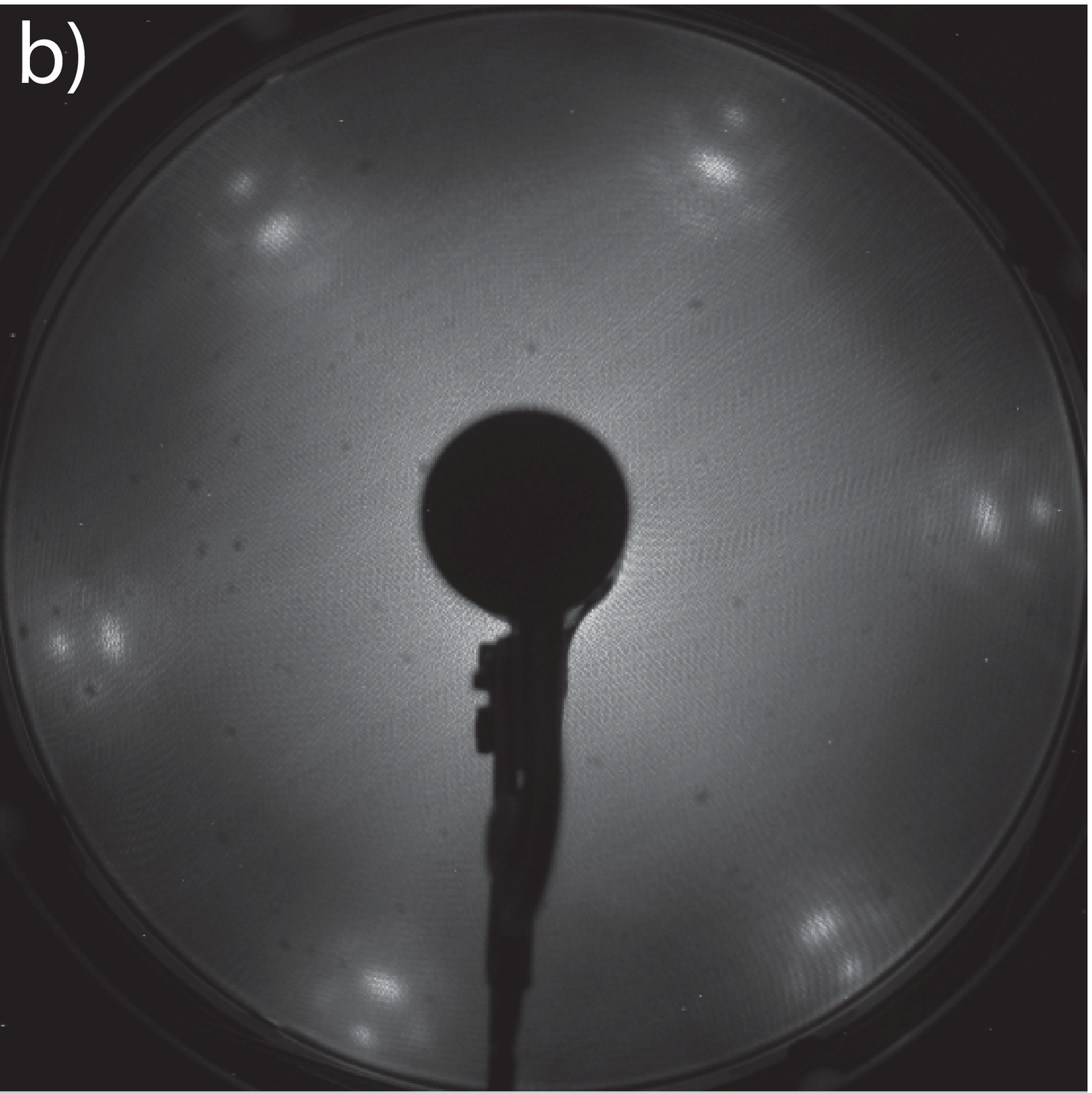}\\
\includegraphics[width=0.2\textwidth]{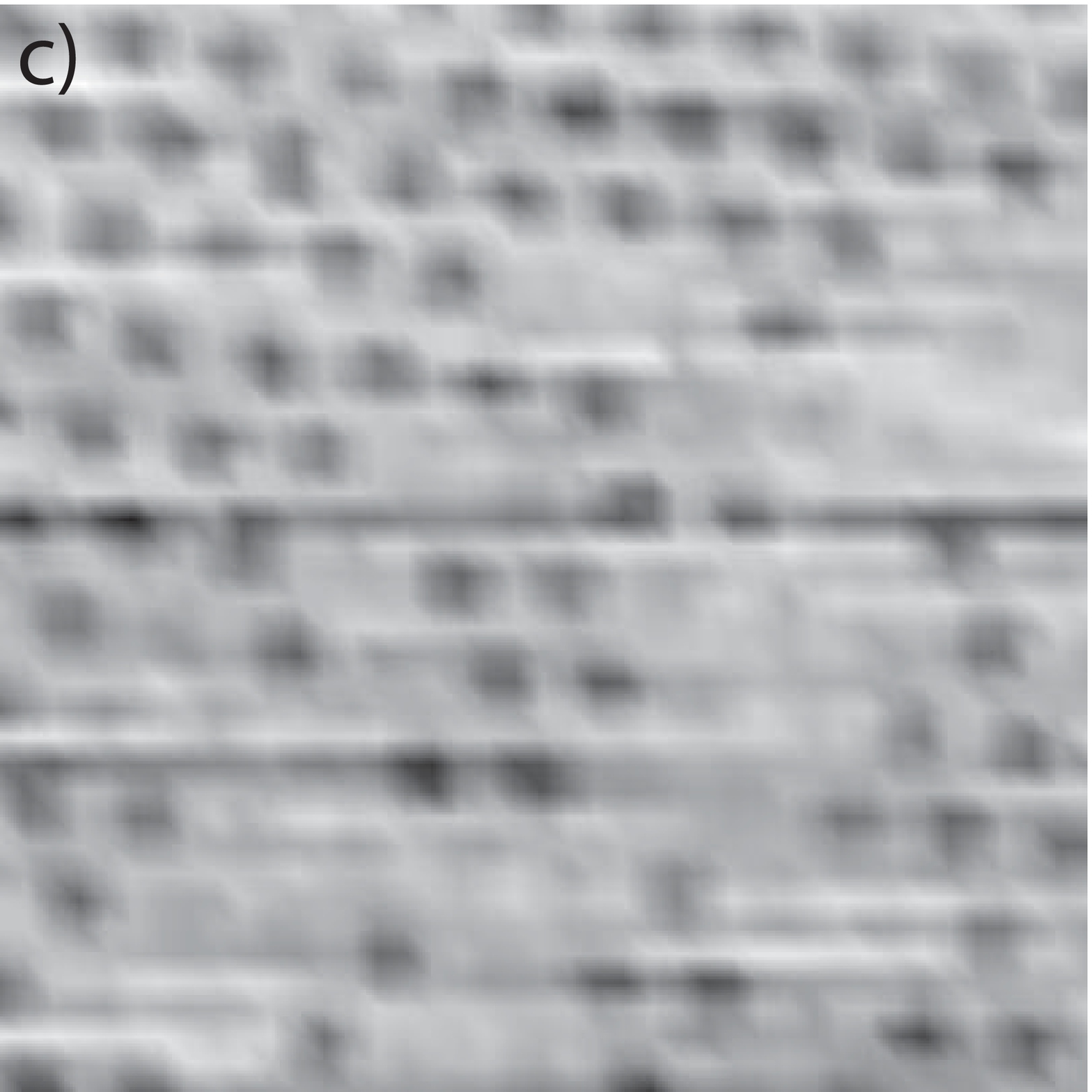}&\includegraphics[width=0.2\textwidth]{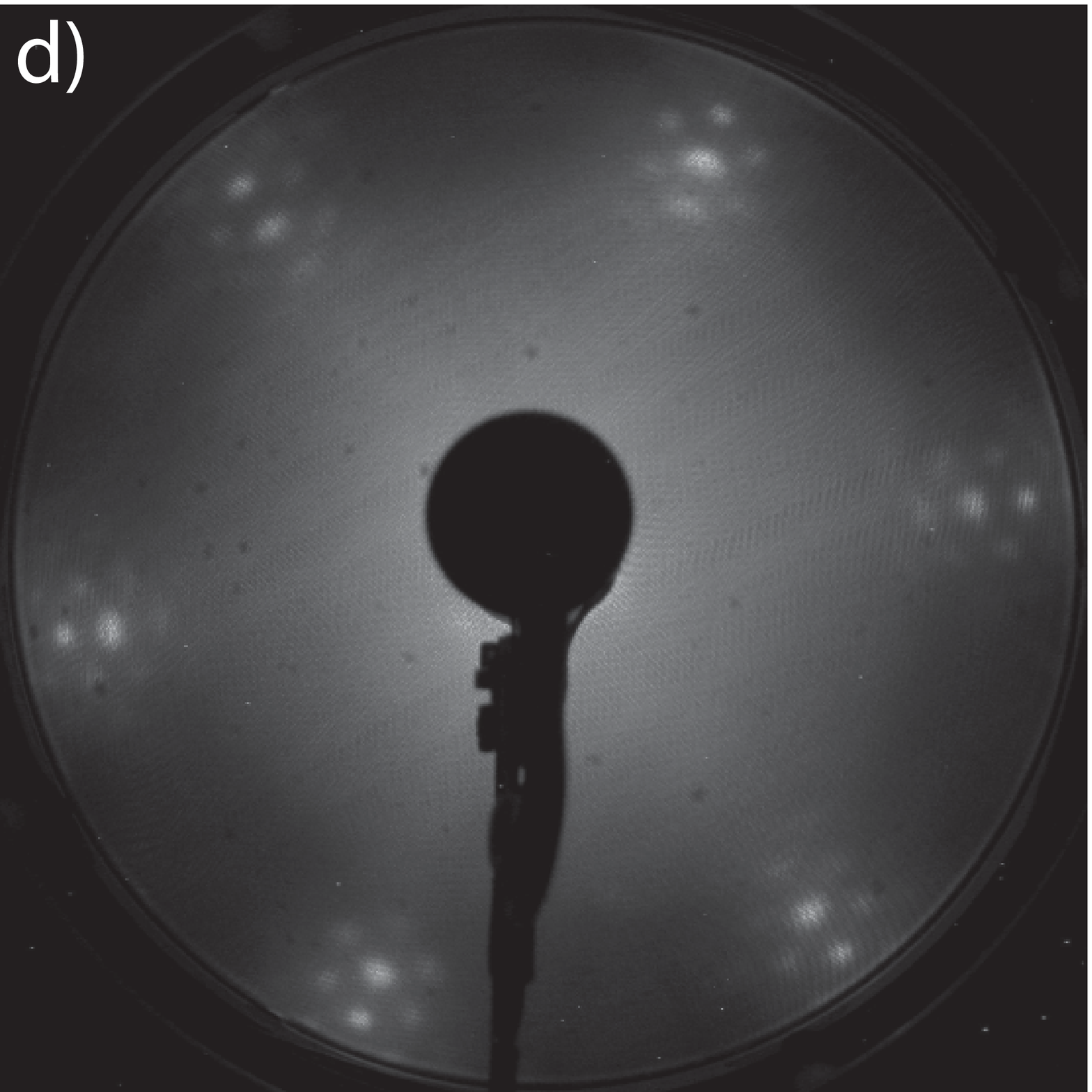}
\end{tabular}
\caption{\label{fig:LEED} \label{fig:20080826m7_closeupWithAg}
a) Closeup STM image of a Ag capped Co island on Cu(111) with b)
corresponding LEED pattern. The observed Moir\'e pattern has a
corrugation of 25 pm and a periodicity of ca.~24 {\AA} taken with the
bias voltage of 0.35 V. c) Closeup STM image of the (9$\times$9) reconstruction
of Ag grown directly on Cu(111) with the bias voltage at 0.23 V and d)
corresponding LEED pattern. The image size is 24$\times$24 nm in both a)
and c). The electron energy used for the LEED images was 65 eV.}
\end{figure}

Finally, we will address the atomic structure of Ag on Co/Cu(111) by
STM and LEED measurements. Figure~\ref{fig:LEED}a
shows an STM image of bilayer thick Co island which subsequently has been capped with Ag. We
observe a Moir\'e pattern with a periodicity of ca.~24 {\AA}
and an apparent corrugation of 25 pm. For the diffraction
experiments we deposited 3 ML Co and subsequently 1.3 ML Ag at room temperature. The LEED image in Figure~\ref{fig:LEED}b show six main spots from the Cu(111) substrate (or the Co islands which grow
pseudomorphically \cite{Figuera1996}). At each main spot there are six
satellites due to the Ag layer. This symmetry is identical to what is seen for a Ag overlayer on
Cu(111) \cite{Bauer1967,Mitchell1974,McMahon1992,Aufray1997,Bendounan2003} where the atomic structure depends on deposition temperature \cite{McMahon1992,Bendounan2003}, but in any case the periodicity is ca.~(9$\times$9) originating from the lattice mismatch between Cu and Ag. For
comparison we have deposited 1.3 ML Ag on Cu(111) and recorded an STM and a
LEED image (shown in Figure~\ref{fig:LEED}c and d).
 Since Co grows pseudomorphically on Cu a similar reconstruction is
expected for Ag grown on Co/Cu(111). This is indeed observed here,
although the Ag-related LEED pattern is more blurry for
Ag/Co/Cu(111) compared to Ag/Cu(111). We believe this is due to edge
effects at Co islands where the periodicity of the Ag Moir\'e pattern
may change and the relatively smaller "coherence areas" for LEED
taken on Ag/Co/Cu(111) than on the much larger Ag areas directly on
Cu(111), since Ag grows almost layer by layer on
Cu(111) \cite{McMahon1992} and the size of Ag domains is therefore
only limited by the terrace size of the Cu(111) crystal. The Moir\'e pattern of Ag on Co/Cu(111) has recently been shown to exhibit interesting electronic properties, where the electronic structure of the Co islands is modulated with the same periodicity as the Ag overlayer Moir\'e superstructure \cite{Bork2009}.

\section{\label{sec:conclusion}Conclusion}
We have studied the growth of a Ag overlayer on Co nanoislands on Cu(111) at room temperature. The atomic structure of Ag on Co/Cu(111) shows a (9$\times$9) periodicity due to the mismatch of lattice parameter. We find that there is a preferred nucleation for Ag on top of Co compared to Cu(111) terrace sites, especially at low Ag doses. This is quite remarkable since the Co islands are all of bilayer height and an accumulation on top of the islands thus involves ascending steps of 2 ML height. We furthermore find that Co islands are either completely free of Ag or almost completely capped with Ag. These nearly perfect multilayers can be expected to act as a model system for magnetic multilayer studies.

%
%
\end{document}